\documentclass{article}%
\usepackage{amsfonts}
\usepackage{aip}
\usepackage{amsmath}
\usepackage{amssymb}
\usepackage{graphicx}%
\providecommand{\U}[1]{\protect\rule{.1in}{.1in}}
\newtheorem{theorem}{Theorem}
\newtheorem{acknowledgement}[theorem]{Acknowledgement}

\begin{document}

\title{On the Consistency of the Solutions of the Space Fractional Schr\"{o}dinger Equation}
\author{Selcuk S. Bayin\\Middle East Technical University\\Institute of Applied Mathematics\\Ankara TURKEY 06800}
\date{\today}
\maketitle

\begin{abstract}
Recently it was pointed out that the solutions found in literature for the
space fractional Schr\"{o}dinger equation in a piecewise manner are wrong,
except the case with the delta potential. We reanalyze this problem and show
that an exact and a proper treatment of the\ relevant integral proves
otherwise. We also discuss effective potential approach and present a free
particle solution for the space and time fractional Schr\"{o}dinger equation
in general coordinates in terms of Fox's H-functions.


\end{abstract}

\section{Introduction}

In a recent article, Jeng et. al. [1] argued that solutions obtained for the
space fractional Schr\"{o}dinger equation in a piecewise fashion are not
valid. Their arguments were based on a contradiction that they observed when
the ground state wave function of the infinite potential well problem is
substituted back into the space fractional Schr\"{o}dinger equation. They
concluded that many exact solutions presented in literature are wrong, except
the one with the delta function potential. This not only casts doubt on the
existing solutions in literature, but also makes it very difficult to find a
meaningful solution. In this manuscript, we show that an exact treatment of
the integral that lead them to inconsistency proves otherwise.

Fractional calculus has found to be extremely useful in the study of anomalous
diffusion. The fastest way to introduce memory effects into an existing
equation is to replace a time derivative with its fractional counterpart. In
this regard, recently we have reconsidered the time fractional Schr\"{o}dinger
equation and its separable solutions [2]. The Space fractional Schr\"{o}dinger
equation naturally incorporates non local effects and was first introduced by
Laskin [3-7]. It is intriguing that the path integral formulation of quantum
mechanics over L\'{e}vy paths, which has many interesting examples in nature,
leads naturally to the space fractional Schr\"{o}dinger equation where the
space derivative is replaced by the fractional Riesz derivative [2,7]. In this
regard, Riesz derivative plays a central role in the fractional generalization
of the Schr\"{o}dinger equation, where solutions for various potentials are
available [3-9]. It should be emphasized that even thought memory and/or non
local effects can be studied via the fractional generalizations of certain
derivatives in physical equations, there could also be\ interesting cases
where fractional derivatives alone are not sufficient [10,11]. For example,
path integral formulations of quantum mechanics over distributions other than
the L\'{e}vy distribution, would not necessarily lead to a fractional
Schr\"{o}dinger equation. Fractional extension of quantum mechanics that
treats coordinates and conjugated momentum equivalently have also been
considered in literature [12].

To understand the effects of fractional derivatives in Schr\"{o}dinger
equation, one could try effective potentials within the framework of quantum
mechanics. In this regard, we have introduced the effective potential approach
for the time fractional Schr\"{o}dinger equation [2], which allowed us to
write the time dependence of the separable solutions, which is a
Mittag-Leffler function with an imaginary argument, as the product of its
purely oscillating and purely decaying parts. In this manuscript, we also
write the effective potential for the space fractional Schr\"{o}dinger
equation and apply it to the infinite well problem. We also present a general
solution for the space and time fractional Schr\"{o}dinger equation in terms
of Fox's H-functions in general coordinates.

\section{Consistency of the Solutions of the Space Fractional Schr\"{o}dinger
Equation}

It is well known that the one dimensional space fractional Schr\"{o}dinger
equation is given as%
\begin{equation}
i\hslash\frac{\partial\Psi}{\partial t}=D_{\alpha}\left(  -\hslash^{2}%
\Delta\right)  ^{\alpha/2}\Psi+V\Psi,
\end{equation}
where $\left(  -\hslash^{2}\Delta\right)  ^{\alpha/2}$ is the quantum Riesz
derivative [7,13]:%
\begin{equation}
\left(  -\hslash^{2}\Delta\right)  ^{\alpha/2}\Psi(x,t)=\frac{1}{2\pi\hslash
}\int_{-\infty}^{+\infty}dp\text{ }e^{ipx/\hslash}\left\vert p\right\vert
^{\alpha}\Phi(p,t),
\end{equation}
and $\Phi(p,t)$ is the Fourier transform of $\Psi(x,t):$%

\begin{equation}
\Phi(p,t)=\int_{-\infty}^{+\infty}dx\text{ }\Psi(x,t)e^{-ipx/\hslash}.
\end{equation}
For the infinite square well potential:%
\begin{equation}
V(x)=\left\{
\begin{tabular}
[c]{ccc}%
$0$ & if & $\left\vert x\right\vert <a$\\
&  & \\
$\infty$ & if & $\left\vert x\right\vert \geqslant a$%
\end{tabular}
\ \ \ \ \ \ \ \ \right.  ,
\end{equation}
the separable solutions are of the form
\begin{equation}
\Psi(x,t)=e^{-iEt/\hslash}\psi(x),
\end{equation}
where $\psi(x)$ satisfies the eigenvalue problem%
\begin{equation}
D_{\alpha}\left(  -\hslash^{2}\Delta\right)  ^{\alpha/2}\psi(x)=E\psi
(x),\text{ }\psi(a)=\psi(-a)=0.
\end{equation}
The general solution is obtained as [4,7,8]%
\begin{equation}
\psi_{n}(x)=\left\{
\begin{tabular}
[c]{ccc}%
$A\cos\left(  \frac{n\pi}{2a}(x+a)\right)  $ & if & $\left\vert x\right\vert
<a$\\
&  & \\
$0$ & if & $\left\vert x\right\vert \geqslant a$%
\end{tabular}
\ \ \ \ \ \ \ \ \right.  ,\text{ }n=1,2,\ldots
\end{equation}
with the energy eigenvalues
\begin{equation}
E_{n}=D_{\alpha}\left(  \frac{\hslash n\pi}{2a}\right)  ^{\alpha}.
\end{equation}
To show the inconsistency of these solutions, Jeng et. al. [1] concentrated on
the ground state with $n=1$:%
\begin{equation}
\psi_{1}(x)=\left\{
\begin{tabular}
[c]{ccc}%
$A\cos\left(  \frac{\pi x}{2a}\right)  $ & if & $\left\vert x\right\vert <a$\\
&  & \\
$0$ & if & $\left\vert x\right\vert \geqslant a$%
\end{tabular}
\ \ \ \ \ \ \ \ \right.  .
\end{equation}
and argued that this solution, albeit satisfying the boundary conditions,
$\psi_{1}(-a)=\psi_{1}(a)=0,$ when substituted back into the space fractional
Schr\"{o}dinger equation [Eq. (6)] leads to a contradiction.

The Fourier transform of $\psi_{1}(x)$ can be found as%
\begin{equation}
\phi_{1}(p)=-A\pi\left(  \frac{\hslash^{2}}{a}\right)  \frac{\cos\left(
ap/\hslash\right)  }{p^{2}-\left(  \pi\hslash/2a\right)  ^{2}},\text{
}\left\vert x\right\vert <a,
\end{equation}
which when substituted into Equation (6), gives $\psi_{1}(x)$ as the integral
\begin{equation}
\psi_{1}(x)=-\frac{AD_{\alpha}}{2E_{1}}\left(  \frac{\hslash}{a}\right)
\int_{-\infty}^{+\infty}dp\text{ }\left(  \frac{2a}{\pi\hslash}\right)
^{2}\frac{\left\vert p\right\vert ^{\alpha}\cos\left(  ap/\hslash\right)
}{\ \left(  2ap/\pi\hslash\right)  ^{2}-1}e^{ipx/\hslash},\text{ }\left\vert
x\right\vert <a.
\end{equation}
Defining a new variable, $q=\frac{2a}{\pi\hslash}p,$ we can also write this as%
\begin{equation}
\psi_{1}(x)=-\frac{AD_{\alpha}}{\pi E_{1}}\left(  \frac{\pi\hslash}%
{2a}\right)  ^{\alpha}\int_{-\infty}^{+\infty}dq\text{ }\frac{\left\vert
q\right\vert ^{\alpha}\cos\left(  \pi q/2\right)  }{\ q^{2}-1}e^{i\pi qx/2a}.
\end{equation}
Without evaluating the above integral, Jeng et. al. [1] argued that the right
hand side, which they wrote as%
\begin{equation}
\psi_{1}(x)=-\frac{AD_{\alpha}}{\pi E_{1}}\left(  \frac{\pi\hslash}%
{2a}\right)  ^{\alpha}2\int_{0}^{+\infty}dq\text{ }\frac{\left\vert
q\right\vert ^{\alpha}\cos\left(  \pi q/2\right)  }{\ q^{2}-1}\cos\left(  \pi
qx/2a\right)  ,
\end{equation}
can not satisfy the boundary conditions as $x\rightarrow\pm a$ that $\psi
_{1}(x)$ satisfies, thus pointing to an inconsistency. However, an exact
evaluation of the integral\ in Equation (12) proves otherwise. Note that the
integral
\begin{equation}
I=\int_{-\infty}^{+\infty}dq\text{ }\frac{\left\vert q\right\vert ^{\alpha
}\cos\left(  \pi q/2\right)  }{\ q^{2}-1}e^{i\pi qx/2a}%
\end{equation}
is a singular integral with poles on the real axis at $q=\pm1,$ which could be
evaluated via analytic continuation as a Cauchy principal value integral [13].

Substituting
\begin{equation}
\cos\left(  \pi q/2\right)  =\frac{1}{2}\left(  e^{i\pi q/2}+e^{-i\pi
q/2}\right)  ,
\end{equation}
we first write $I$ as two integrals:%
\begin{align}
I  &  =\frac{1}{2}\int_{-\infty}^{+\infty}dq\text{ }\frac{\left\vert
q\right\vert ^{\alpha}\ e^{i(x/a+1)(\pi q/2)}}{\ (q+1)(q-1)}\ +\frac{1}{2}%
\int_{-\infty}^{+\infty}dq\text{ }\frac{\left\vert q\right\vert ^{\alpha
}\ e^{i(x/a-1)(\pi q/2)}}{\ (q+1)(q-1)}\\
&  =I_{1}+I_{2}.
\end{align}
For the first integral, $I_{1},$ we close the contour in the upper half
$q-$plane with a semicircular path of radius $R$ and then go around the poles
on the real axis in the upper $q-$ plane, with small semicircular paths of
radius $\delta$. In the limit as $R\rightarrow\infty$, via Jordan's lemma, the
contribution over the large semicircle vanishes, thus allowing us to evaluate
the value of this integral as a Cauchy principal value in the limit
$\delta\rightarrow0$ as [13]%
\begin{equation}
PV(I_{1})=i\pi\left(  \frac{i}{2}\cos\frac{x\pi}{2a}\right)  .
\end{equation}
Similarly for $I_{2},$ we close the contour this time in the lower $q-$plane
and circle around the poles in the lower half $q-$plane to obtain%
\begin{equation}
PV(I_{2})=-i\pi\left(  -\frac{i}{2}\cos\frac{x\pi}{2a}\right)  .
\end{equation}
Adding these we obtain the Cauchy principal value of $I$ as%
\begin{equation}
PV(I)=-\pi\cos\frac{x\pi}{2a},
\end{equation}
which on the contrary to Jeng et. al. [1] vanishes at the boundary as
$x\rightarrow\pm a$. When this is substituted back into Equation (12) yields%
\begin{equation}
\psi_{1}(x)=\frac{AD_{\alpha}}{E_{1}}\left(  \frac{\pi\hslash}{2a}\right)
^{\alpha}\cos\frac{x\pi}{2a},\text{ }\left\vert x\right\vert <a.
\end{equation}
Since the eigenvalue for the ground state is%
\begin{equation}
E_{1}=D_{\alpha}\left(  \frac{\pi\hslash}{2a}\right)  ^{\alpha},
\end{equation}
we again obtain the ground state wave function [Eq. (9)]:%
\begin{equation}
\psi_{1}(x)=A\cos\frac{x\pi}{2a},\text{ }\left\vert x\right\vert <a,
\end{equation}
which satisfies the boundary conditions, hence no inconsistency.

The proof for general $n$ also follows along the same lines. For example, for
the odd values of $n,$ Equation (12) is replaced by%
\begin{equation}
\psi_{n}(x)=-\frac{AD_{\alpha}}{\pi E_{n}}\sin\left(  \frac{n\pi}{2}\right)
\left(  \frac{n\pi\hslash}{2a}\right)  ^{\alpha}\int_{-\infty}^{+\infty
}dq\text{ }\frac{\left\vert q\right\vert ^{\alpha}\cos\left(  n\pi q/2\right)
}{\ q^{2}-1}e^{i(n\pi x/2a)q}.
\end{equation}
Now the Cauchy principal value of the needed integral:%
\begin{equation}
I=\int_{-\infty}^{+\infty}dq\text{ }\frac{\left\vert q\right\vert ^{\alpha
}\cos\left(  n\pi q/2\right)  }{\ q^{2}-1}e^{i(n\pi x/2a)q},
\end{equation}
is found as%
\begin{equation}
PV(I)=-\pi\sin\left(  \frac{n\pi}{2}\right)  \cos\left(  \frac{n\pi x}%
{2a}\right)  ,
\end{equation}
which when substituted back into Equation (24) yields%
\begin{align}
\psi_{n}(x)  &  =\frac{AD_{\alpha}}{\pi E_{n}}\sin^{2}\left(  \frac{n\pi}%
{2}\right)  \left(  \frac{n\pi\hslash}{2a}\right)  ^{\alpha}\cos\left(
\frac{n\pi x}{2a}\right)  ,\\
&  =A\cos\left(  \frac{n\pi x}{2a}\right)  ,\text{ }n=1,3,5,\ldots\text{.}%
\end{align}
Again, no contradiction. The proof for the even values of $n$ is similar.

\section{Effective Potential for the Space Fractional Schr\"{o}dinger
Equation}

For the separable solutions of the free space fractional Schr\"{o}dinger
equation we use Laskin's notation:
\begin{equation}
i\hbar\frac{\partial}{\partial t}\Psi(x,t)=-D_{\beta}(\hbar\nabla)^{\beta}%
\Psi(x,t),
\end{equation}
where $\nabla^{\beta}$ is the Riesz derivative and
\begin{equation}
\left(  -\hslash^{2}\Delta\right)  ^{\alpha/2}\Psi(x,t)=-(\hbar\nabla)^{\beta
}\Psi(x,t).
\end{equation}
For a separable solution, $\Psi(x,t)=T(t)X(x),$ found by solving the free
space fractional Schr\"{o}dinger equation we can define an effective potential
in terms of the Schr\"{o}dinger equation as $\ $
\begin{equation}
V_{eff.}(x)=\frac{\hbar^{2}}{2m}\frac{1}{X(x)}\left(  \frac{d^{2}X(x)}{dx^{2}%
}\right)  -\frac{D_{\beta}(\hbar\nabla)^{\beta}X(x)}{X(x)}\ .
\end{equation}
For separable solutions, the free space fractional Schr\"{o}dinger equation
reduces to%

\begin{align}
i\hbar\frac{dT(t)}{dt}  &  =E_{n}T(t),\\
-D_{\beta}(\hbar\nabla)^{\beta}X(x)  &  =E_{n}X(x),
\end{align}
where the energy eigenvalues, $E_{n},$ come from the solution of Equation (33)
with the appropriate boundary conditions and, the time dependence is given as
\begin{equation}
T(t)=e^{-i(E_{n}/\hbar)t}.
\end{equation}
Since in the presence of a potential, $V(x),$ the corresponding fractional
Hamiltonian:
\begin{equation}
H_{\beta}=-D_{\beta}(\hbar\nabla)^{\beta}+V(x),
\end{equation}
is hermitian, the energy eigenvalues are real [7]. Thus, we can also write
\begin{equation}
V_{eff.}(x)=\frac{\hbar^{2}}{2m}\frac{1}{X(x)}\left(  \frac{d^{2}X(x)}{dx^{2}%
}\right)  +E_{n}.
\end{equation}
The equivalent Schr\"{o}dinger equations are now written as
\begin{align}
i\hbar\frac{dT(t)}{dt}  &  =E_{n}T(t),\\
-\frac{\hbar^{2}}{2m}\frac{1}{X(x)}\left(  \frac{d^{2}X(x)}{dx^{2}}\right)
+V_{eff.}(x)  &  =E_{n},
\end{align}
which are to be solved with the same boundary conditions.

\subsection{Effective Potential for the Infinite Well}

For the box problem, using the normalization condition%
\begin{equation}
\int_{-a}^{a}dx\left\vert \Psi(x,t)\right\vert ^{2}=1,
\end{equation}
the complete wave function is given as
\begin{equation}
\Psi(x,t)=\left\{
\begin{tabular}
[c]{lll}%
$\frac{1}{\sqrt{a}}e^{-(i/\hbar)E_{n}t}\sin\dfrac{n\pi}{2a}(x+a)$ & $,$ &
$\left\vert x\right\vert <a$\\
&  & \\
$0$ & $,$ & $\left\vert x\right\vert \geq a$%
\end{tabular}
\ \ \ \ \ \ \ \ \right.  \text{ , }n=1,2,\ldots\text{ ,}%
\end{equation}
where
\begin{equation}
E_{n}=D_{\beta}\left(  \frac{n\pi\hbar}{2a}\right)  ^{\beta}\text{ and }%
\Psi(-a,t)=\Psi(a,t)=0.
\end{equation}
The effective potential is now written as
\begin{equation}
V_{eff.}(x)=D_{\beta}\left(  \frac{n\pi\hbar}{2a}\right)  ^{\beta}-\frac
{\hbar^{2}}{2m}\left(  \frac{n\pi}{2a}\right)  ^{2}.
\end{equation}
In other words, the effective potential shifts the energy levels by a constant
as a function of the quantum number $n$.

\section{Free Particle Solution of the Space and Time Fractional
Schr\"{o}dinger Equation}

We now consider the free Schr\"{o}dinger equation with both time and space
fractional derivatives [2]:
\begin{equation}
\frac{\partial^{\alpha}}{\partial t^{\alpha}}\Psi(x,t)=\frac{i^{\alpha}}%
{\hbar}\check{D}_{\alpha,\beta}\left(  \hbar\nabla\right)  ^{\beta}%
\Psi(x,t),\text{ }0<\alpha<1,\text{ }1<\beta<2,
\end{equation}
where $\frac{\partial^{\alpha}}{\partial t^{\alpha}}$ is the Caputo derivative
(Appendix A). Performing a Wick rotation gives the Bloch equation:
\begin{equation}
\frac{\partial^{\alpha}}{\partial t^{\alpha}}\Psi(x,t)=\frac{1}{\hbar}%
\check{D}_{\alpha,\beta}\hbar^{\beta}\nabla_{x}^{\beta}\Psi(x,t),\text{
}0<\alpha<1,\text{ }1<\beta<2,
\end{equation}
where $\check{D}_{1,2}=1/2m.$ Using the boundary conditions
\begin{equation}
\Psi\left(  x,0\right)  =\delta(x)\text{ \ and \ }\lim_{x\rightarrow\pm\infty
}\Psi(x,t)\rightarrow0,
\end{equation}
we take the Laplace transform with respect to time and the Fourier transform
with respect to space to obtain the Fourier-Laplace transform of the solution
[14]. Finding the inverse transform and then performing an inverse Wick
rotation, yields the wave function in integral form as
\begin{equation}
\Psi(x,t)=\frac{\Psi_{0}}{2\pi}\int_{-\infty}^{+\infty}e^{-ikx}E_{\alpha
}\left(  -\frac{i^{\alpha}}{\hbar}\check{D}_{\alpha,\beta}\hbar^{\beta
}k^{\beta}t^{\alpha}\right)  \text{ }dk,
\end{equation}
where $E_{\alpha}(z)$ is the Mittag-Leffler function [2].\ We can also write
$\Psi(x,t)$ as
\begin{equation}
\Psi(x,t)=\frac{\Psi_{0}}{\pi}\int_{0}^{+\infty}\cos kx\text{ }E_{\alpha
}\left(  -\frac{i^{\alpha}}{\hbar}\check{D}_{\alpha,\beta}\hbar^{\beta
}k^{\beta}t^{\alpha}\right)  \text{ }dk.
\end{equation}
This wave function satisfies both the time and the space fractional
Schr\"{o}dinger equation [Eq. (43)]

In terms of $H-$functions (Appendix B) Equation (47) can be written as
\begin{equation}
\Psi(x,t)=\frac{\Psi_{0}}{\pi}\int_{0}^{+\infty}\cos kx\text{ }H_{1,2}%
^{1,1}\left(  \left.  \frac{i^{\alpha}}{\hbar}\check{D}_{\alpha,\beta}%
\hbar^{\beta}k^{\beta}t^{\alpha}\right\vert _{(0,1),(0,\alpha)}^{(0,1)}%
\right)  \text{ }dk,
\end{equation}
which can be integrated by using the properties of the H$-$functions [15]
as$\ $%
\begin{equation}
\Psi(x,t)=\frac{\Psi_{0}}{\sqrt{\pi}\left\vert x\right\vert }H_{3,2}%
^{1,2}\left(  \left.  \frac{i^{\alpha}}{\hbar}\check{D}_{\alpha,\beta}%
\hbar^{\beta}t^{\alpha}\left(  \frac{2}{\left\vert x\right\vert }\right)
^{\beta}\right\vert _{(0,1),(0,\alpha)}^{(1/2,\beta/2),(0,1),(0,\beta
/2)}\right)  .
\end{equation}
To determine $\Psi_{0}$ the final wave function has to be normalized as
\begin{equation}
\int\left\vert \Psi(x,0)\right\vert ^{2}dx=1.
\end{equation}

Case I:

For $\beta=2,$ this solution becomes the free particle solution of the time
fractional Schr\"{o}dinger equation as
\begin{equation}
\Psi(x,t)=\frac{\Psi_{0}}{\sqrt{\pi}\left\vert x\right\vert }H_{3,2}%
^{1,2}\left(  \left.  \frac{4i^{\alpha}D_{\alpha}t^{\alpha}}{\left\vert
x\right\vert ^{2}}\right\vert _{(0,1),(0,\alpha)}^{(1/2,1),(0,1),(0,1)}%
\right)  ,\text{ }D_{\alpha}=\check{D}_{\alpha,2}\hbar.
\end{equation}
This can also be shown to be equal to
\begin{equation}
\Psi(x,t)=\frac{\Psi_{0}}{\sqrt{\pi}\left\vert x\right\vert }H_{1,2}%
^{2,0}\left(  \left.  \frac{\left\vert x\right\vert ^{2}}{4i^{\alpha}%
D_{\alpha}t^{\alpha}}\right\vert _{(1/2,1),(1,1)}^{(1,\alpha)}\right)  ,
\end{equation}
or to
\begin{equation}
\Psi(x,t)=\frac{\Psi_{0}}{\left\vert x\right\vert }H_{1,1}^{1,0}\left(
\left.  \frac{\left\vert x\right\vert ^{2}}{i^{\alpha}D_{\alpha}t^{\alpha}%
}\right\vert _{(1,2)}^{(1,\alpha)}\right)  .
\end{equation}
In the limit as $\alpha\rightarrow1,$ this becomes
\begin{equation}
\Psi(x,t)=\frac{\Psi_{0}}{(4\pi iD_{1}t)^{1/2}}\exp\left(  \ -\frac{\left\vert
x\right\vert ^{2}}{4iD_{1}t}\right)  ,
\end{equation}
where $D_{1}=\hbar/2m.$

Case II:

When $\alpha=1,$ we obtain the wave function as
\begin{equation}
\Psi(x,t)=\frac{\Psi_{0}}{\sqrt{\pi}\left\vert x\right\vert }H_{3,2}%
^{1,2}\left(  \left.  \frac{i}{\hbar}D_{\beta}\hbar^{\beta}t\left(  \frac
{2}{\left\vert x\right\vert }\right)  ^{\beta}\text{ }\right\vert
_{(0,1),(0,1)}^{(1/2,\beta/2),(0,1),(0,\beta/2)}\right)  ,
\end{equation}
which satisfies
\begin{equation}
i\ \hbar\frac{\partial}{\partial t}\Psi(x,t)=-D_{\beta}\hbar^{\beta}%
R_{x}^{\beta}\Psi(x,t),\text{ }1<\beta<2,
\end{equation}
where
\begin{equation}
\check{D}_{1,\beta}=D_{\beta}.
\end{equation}
This solution was also given by [8] in the form:
\begin{equation}
\Psi(x,t)=\frac{\pi\Psi_{0}}{\beta\left\vert x\right\vert }H_{2,2}%
^{1,1}\left(  \left.  \frac{1}{\hbar}\left(  \frac{\hbar}{iD_{\beta}t}\right)
^{1/\beta}\left\vert x\right\vert \right\vert _{(1,1),(1,1/2)}^{(1,1/\beta
),(1,1/2)}\right)  .
\end{equation}

\section{Conclusions}

Successful applications of fractional calculus to anomalous diffusion
attracted researchers from many different branches of science and engineering
into this intriguing branch of mathematics [15-19]. Applications of fractional
calculus usually starts by replacing certain derivatives in the evolution or
the transport equations with their fractional counterparts. In general,
replacing a time derivative with its fractional counterpart incorporates
memory effects into the system and makes the process non Markovian, while a
replacement of a space derivative introduces global or non local effects.

In 2000, Laskin [3-7] introduced the path integral formulation of quantum
mechanics over L\'{e}vy paths and showed that the corresponding equation of
motion is the space fractional Schr\"{o}dinger equation. For the infinite
potential well problem, the complete wave function is given as [7,8] \
\begin{equation}
\Psi(x,t)=\left\{
\begin{tabular}
[c]{lll}%
$\dfrac{e^{-(i/\hbar)E_{n}t}}{\sqrt{a}}\sin\dfrac{n\pi}{2a}(x+a)$ & $,$ &
$\left\vert x\right\vert <a$\\
&  & \\
$0$ & $,$ & $\left\vert x\right\vert \geq a$%
\end{tabular}
\ \ \ \ \ \ \ \ \ \ \ \ \ \right.  \text{ , }n=1,2,\ldots\text{ .}%
\end{equation}
Since the energy operator is now given as $i\hbar\dfrac{d}{dt},$ the energy
eigenvalues are given as%
\begin{equation}
E_{n}=D_{\beta}\left(  \frac{\pi n\hbar}{2a}\right)  ^{\beta},\text{
}n=1,2,\ldots.
\end{equation}
Note that for the space fractional Schr\"{o}dinger equation, the Hamiltonian,
$H=-D_{\beta}\left(  \hbar\frac{\partial}{\partial x}\right)  ^{\beta}+V(x),$
is hermitian. An alternate approach to space fractional Schr\"{o}dinger
equation is given by Herrmann [12].

It is true that the Riesz derivative requires a knowledge of the wave function
over the entire space. For the infinite well problem, the system is confined
to the region $\left\vert x\right\vert \leq a$ with \ $\Psi(x,t)=0$ for
$\left\vert x\right\vert \geq a.$ Since the solution for $\left\vert
x\right\vert <a,$ also satisfies the boundary conditions as $x\rightarrow\pm
a,$ the solution inside the well is consistent with the outside. In this
regard, it is not true that the above solution ignores non locality and is
valid only for $\beta\approx1$ [1,12 pg. 107].

In general, in the presence of a potential, $V(x),$ the time independent space
fractional Schr\"{o}dinger equation is given as%
\begin{equation}
D_{\alpha}\left(  -\hslash^{2}\Delta\right)  ^{\alpha/2}\psi(x;E,\alpha
)+V(x)\psi(x;E,\alpha)=E\psi(x;E,\alpha),
\end{equation}
which is to be solved with the appropriate boundary conditions. The Fourier
transform of Equation (61) yields a dispersion relation of the form
$\phi(p;E,\alpha)$, where the wave function is given as
\begin{equation}
\psi(x;E,\alpha)=\frac{1}{2\pi\hslash}\int_{-\infty}^{+\infty}dp\text{ }%
\phi(p;E,\alpha)e^{ipx/\hslash}.
\end{equation}
Since quantization follows from the boundary conditions that $\psi
(x;E,\alpha)$ satisfies, granted that the above integral exists, solutions
satisfying the given boundary conditions can not lead to an inconsistency [7-9].

After showing the consistency of the solutions of the space fractional
Schr\"{o}dinger equation for the one dimensional infinite square well, we have
also given the effective potential for the space fractional Schr\"{o}dinger
equation. This is the potential that the Schr\"{o}dinger equation with the
same boundary conditions will yield the same wave function as the fractional
case. For the space and time fractional Schr\"{o}dinger equation, we give the
free particle solution in terms of $H-$functions, which in the appropriate
limits reproduces the previous solutions. The solutions we give for the free
particle are in general coordinates.

\appendix{APPENDIX}

\section{Basic Definitions of the Fractional Derivatives and Integrals}

The Caputo definition of fractional derivative is given as%
\begin{equation}
\left[  \frac{d^{q}f(t)}{dt^{q}}\right]  _{C}=\frac{1}{\Gamma(1-q)}\int
_{0}^{t}\left(  \frac{df(\tau)}{d\tau}\right)  \frac{d\tau}{(t-\tau)^{q}%
},0<q<1,
\end{equation}
Laplace transform of the Caputo derivative is
\begin{equation}
\pounds \left\{  _{0}^{C}\mathbf{D}_{t}^{q}f(t)\right\}  =s^{q}\widetilde
{f}(s)-\sum_{k=0}^{n-1}s^{q-k-1}\left.  \frac{d^{k}f(t)}{dt^{k}}\right\vert
_{t=0},\text{ }n-1<q\leq n,
\end{equation}
where $_{0}\mathbf{D}_{t}^{q}f(t)\equiv\dfrac{d^{q}f}{dt^{q}}.$

In writing the space fractional diffusion equation or the space fractional
Schr\"{o}dinger equation, we use the Riesz derivative which is defined with
respect to its Fourier transform:
\begin{equation}
\mathcal{F}\left\{  \mathbf{R}_{x}^{q}f(x)\right\}  =-\left\vert
\omega\right\vert ^{q}g(\omega),\text{ }0<q<2.
\end{equation}
This yields the Riesz derivative as
\begin{equation}
\mathbf{R}_{x}^{q}f(x)=-\frac{1}{2\pi}\int_{-\infty}^{+\infty}\left\vert
\omega\right\vert ^{q}g(\omega)e^{i\omega x}d\omega,
\end{equation}
where $g(\omega)$ is the Fourier transform of $f(x)$. Note that
\begin{equation}
\mathbf{R}_{x}^{2}f(x)=\frac{d^{2}}{dx^{2}}f(x).
\end{equation}

\section{Fox's H-Function}

In 1961 Fox introduced the $H-$function, which gives a general way of
expressing a wide class of functions encountered in applied mathematics.
$H-$function provides an elegant and an efficient formalism to handle problems
in fractional calculus. Fox's $H-$function is a generalization of the Meijer's
$G-$function and is defined with respect to a Mellin-Barnes type integral
[15]:
\begin{align}
H_{p,q}^{m,n}(z)  &  =H_{p,q}^{m,n}\left(  \left.  z\right\vert _{(b_{q}%
,B_{q})}^{(a_{p},A_{p})}\right)  =H_{p,q}^{m,n}\left(  \left.  z\right\vert
_{(b_{1},B_{1}),\ldots,(b_{q},B_{q})}^{(a_{1},A_{1}),\ldots,(a_{p},A_{p}%
)}\right) \\
&  =\frac{1}{2\pi i}\int_{C}h(s)z^{-s}ds,
\end{align}
where
\begin{equation}
h(s)=\frac{\prod_{j=1}^{m}\Gamma(b_{j}+B_{j}s)\prod_{j=1}^{n}\Gamma
(1-a_{j}-A_{j}s)\ }{\prod_{j=n+1}^{p}\Gamma(a_{j}+A_{j}s)\prod_{j=m+1}%
^{q}\Gamma(1-b_{j}-B_{j}s)\ },
\end{equation}
$m,n,p,q$ are positive integers satisfying $0\leq n\leq p,$ $1\leq m\leq q,$
and empty products are taken as unity. Also, $A_{j},$ $j=1,\ldots,p,$ and
$B_{j},$ $j=1,\ldots,q,$ are real positive numbers, and $a_{j},$
$j=1,\ldots,p,$ and $b_{j},$ $j=1,\ldots,q,$ are in general complex numbers
satisfying
\begin{equation}
A_{j}(b_{h}+\nu)\neq B_{h}(a_{j}-\lambda-1)\text{ for }\nu,\lambda
=0,1,\ldots;\text{ }h=1,\ldots,m,\text{ }j=1,\ldots,n.
\end{equation}
The contour $C$ is such that the poles of $\Gamma(b_{j}+B_{j}s),$
$j=1,\ldots,m,$ are separated from the poles of $\Gamma(1-a_{j}-A_{j}s),$
$j=1,\ldots,n$ such that the poles of $\Gamma(b_{j}+B_{j}s)$ lie to the left
of $C$, while the poles of $\Gamma(1-a_{j}-A_{j}s)$ are to the right of $C.$
The poles of the integrand are assumed to be simple. The $H-$function is an
analytic function of $z$ for every $\left\vert z\right\vert \neq0$ when
$\mu>0$ and for $0<\left\vert z\right\vert <1/\beta$ when $\mu=0$, where $\mu$
and $\beta$ are defined as
\begin{equation}
\mu=\sum_{j=1}^{q}B_{j}-\sum_{j=1}^{p}A_{j}\
\end{equation}
and
\begin{equation}
\beta=\prod_{j=1}^{p}A_{j}^{A_{j}}\prod_{j=1}^{q}B_{j}^{-B_{j}}.
\end{equation}

Fox's $H-$function is very useful in the study of stochastic processes and in
solving fractional diffusion equations (Gl\"{o}ckle and Nonnenmacher [20],
West and Grigolini [21]). For example, the following useful formula for the
Riemann-Liouville fractional derivative of the $H-$function:
\begin{equation}
_{0}^{R-L}\mathbf{D}_{z}^{\beta}\left[  z^{a}H_{p,q}^{m,n}\left(  \left.
(cz)^{b}\right\vert _{(b_{j},B_{j})}^{(a_{j},A_{j})}\right)  \right]
=z^{a-\beta}H_{p+1,q+1}^{m,n+1}\left(  \left.  (cz)^{b}\right\vert
_{(b_{j},B_{j}),(\beta-a,b)}^{(-a,b),(a_{j},A_{j})}\right)  ,
\end{equation}
where $a,b>0$ and $a+b\min(b_{j}/B_{j})>-1,$ $1\leq j\leq m,$ can be used to
find solutions to the fractional diffusion equation by tuning the indices to
appropriate values. Similarly, the Laplace transform of the $H-$function can
be obtained by using the formula [15]
\begin{equation}
\pounds \left\{  x^{\rho-1}H_{p,q+1}^{m,n}\left(  \left.  ax^{\sigma
}\right\vert _{(b_{q},B_{q}),(1-\rho,\sigma)}^{(a_{p},A_{p})}\right)
\right\}  =s^{-\rho}H_{p,q}^{m,n}\left(  \left.  as^{-\sigma}\right\vert
_{(b_{q},B_{q})}^{(a_{p},A_{p})}\right)  ,
\end{equation}
where the inverse transform is given as
\begin{equation}
\pounds ^{-1}\left\{  s^{-\rho}H_{p,q}^{m,n}\left(  \left.  as^{\sigma
}\right\vert _{(b_{q},B_{q})}^{(a_{p},A_{p})}\right)  \right\}  =x^{\rho
-1}H_{p+1,q}^{m,n}\left(  \left.  ax^{-\sigma}\right\vert _{(b_{q},B_{q}%
)}^{(a_{p},A_{p}),(\rho,\sigma)}\right)  ,
\end{equation}
where
\begin{equation}
\rho,\alpha,s\text{ }\in\mathbb{C},\text{ }\operatorname{Re}(s)>0,\text{
}\sigma>0
\end{equation}
and
\begin{equation}
\operatorname{Re}(\rho)+\sigma\max_{1\leq i\leq n}\left[  \frac{1}{A_{i}%
}+\frac{\operatorname{Re}(a_{i})}{A_{i}}\right]  >0,\text{ }\left\vert \arg
a\right\vert <\frac{\pi\theta}{2},\text{ }\theta=\alpha-\sigma.
\end{equation}

\begin{acknowledgement}
I would like to thank Jean Krisch and Nick Laskin for valuable comments.$\ $
\end{acknowledgement}

\end{document}